\documentclass[a4paper]{jpconf}
\usepackage{graphicx}
\usepackage{footmisc}
\renewcommand{\thefootnote}{\fnsymbol{footnote}}
\begin{document}
\title{Global extraction of the parton-to-kaon fragmentation functions at NLO in QCD}

\author{R. J. Hern\'andez-Pinto$^1$\footnote[1]{\hspace{1mm}Speaker}, M. Epele$^2$, D. de Florian$^3$, R. Sassot$^4$ and M. Stratmann$^5$}

\address{
$^1$Facultad de Ciencias F\'isico-Matem\'aticas, Universidad Aut\'onoma de Sinaloa, Ciudad Universitaria, CP 80000, Culiac\'an, Sinaloa, M\'exico, \\
$^2$
Instituto de F\'isica La Plata, UNLP, CONICET Departamento de F\'isica, Facultad de Ciencias Exactas, Universidad de La Plata, C.C. 69, La Plata, Argentina, \\
$^3$
International Center for Advanced Studies (ICAS), UNSAM, Campus Miguelete, 25 de Mayo y Francia, (1650) Buenos Aires, Argentina, \\
$^4$
Departamento de F\'isica and IFIBA, Facultad de Ciencias Exactas y Naturales, Universidad de Buenos Aires, Ciudad Universitaria, Pabell\'on 1 (1428) Buenos Aires, Argentina, \\
$^5$
Institute for Theoretical Physics, University of T\"ubingen, Auf der Morgenstelle 14, 72076 T\"ubingen, Germany}

\ead{roger@uas.edu.mx}

\begin{abstract}
In this document, we present the global QCD analysis of parton-to-kaon fragmentation functions at next-to-leading order accuracy using the latest experimental information on single-inclusive kaon production in electron-positron annihilation, lepton-nucleon deep-inelastic scattering, and proton-proton collisions. An extended analysis of this work can be found in Ref.~\cite{DSS17}.
\end{abstract}

\section{Introduction}
Parton-to-hadron fragmentation functions (FFs) parametrize how quarks and gluons that are produced
in hard interactions at high energies confine themselves into hadrons measured and identified in experiment.
This information is beyond the reach of perturbative Quantum Chromodynamics (pQCD) and must therefore
be inferred from the wealth of data on identified hadron production under the theoretical assumption that the
relevant non-perturbative dynamics of FFs factorizes in a universal way from the calculable hard partonic cross
sections up to small corrections which can be usually neglected.
A precise knowledge on FFs is vital for the quantitative description of a wide variety of hard scattering
processes designed to probe the spin and flavor structure of nucleons and nuclear matter and their interpretation
at the most elementary and fundamental level. 
In this review, we present the basic features of the updated parton-to-kaon FFs.

\section{Functional form and fit parameters}
Similar to the updated pion FFs~\cite{DSS14}, we found that the functional form adopted in Ref.~\cite{DSS07} is still flexible enough to accommodate also
the wealth of new experimental information included in the present fit. Specifically, we parametrize
the hadronization of a parton of flavor $i$ into a positively charged kaon $K^+$ at an initial scale 
of $Q_0 = 1$ GeV as
\begin{eqnarray}
D^{K^+}_i (z, Q_0) = \frac{N_i z^{\alpha_i} (1 - z)^{\beta_i}
[1 + \gamma_i(1 - z)^{\delta_i}]}{
B[2 + \alpha_i
, \beta_i + 1] + \gamma_i B[2 + \alpha_i
, \beta_i + \delta_i + 1]} .
\end{eqnarray}
Here, $B[a, b]$ denotes the Euler Beta-function, and the $N_i$ are chosen in such a way that they represent the
contribution of $z D^{K^+}_i$ to the momentum sum rule.
The corresponding FFs for negatively charged kaons are obtained by charge conjugation symmetry.
In order to reduce the number of parameters in the fit we impose, without changing the total $\chi^2$, the following constraints:
$
\beta_g = \beta_{\bar{u}}, \hspace{1mm} 
1\gamma_{s+\bar{s}}\simeq\gamma_{\bar{u}}, \hspace{1mm} 
\delta_{s+\bar{s}}\simeq\delta_{\bar{u}}, \hspace{1mm} 
\gamma_{c+\bar{c}} = \gamma_{b+\bar{b}}=0, \hspace{1mm} 
\alpha_{c+\bar{c}} = \alpha_{b+\bar{b}}=0. \hspace{1mm} 
$
In total we now have 20 free fit parameters describing
our updated FFs for quarks, antiquarks, and gluons into a positively charged kaon. These parameters are then determined from
data by a standard $\chi^2$-minimization procedure that includes a $\chi^2$-penalty from computing the optimum relative
normalization of each experimental set of data analytically
as was outlined in the DSS 14~\cite{DSS14}.

\section{Data selection}
In addition to the data sets already used in Ref.~\cite{DSS07} for SIA, we now introduce the new results from BaBar~\cite{Babar} and Belle~\cite{Belle} in SIA at a c.m.s. energy of $\sqrt{s}=10.5$ GeV. Both sets are very precise and reach all the way up to kaon momentum fractions $z$ close to one.
We analyze both sets with $n_f = 4$ active, massless flavors using the standard expression for the SIA cross section at NLO accuracy. As is customary, we limit ourselves to data with $z \geq 0.1$ 
to avoid any potential impact from kinematical regions where hadron mass corrections, proportional to $ M_K/(s \,z^2)$, might become of any importance.  For all previous SIA data, taken at higher $\sqrt{s}$, we use $n_f=5$ and also $z \geq 0.1$.

In case of SIDIS, we replace the preliminary multiplicity data from Hermes by their final results~\cite{Hermes}. More specifically, we use the data for charged kaon multiplicities in four bins of $z$ as a function of both momentum transfer $Q^2$ and the target nucleon's (proton or deuteron) momentum fraction $x$. The kinematical ranges of average
values of $Q^2$ and $x$ covered by these data are from about 1.1 GeV$^2$ to 7.4 GeV$^2$ and 0.064 to 0.277, respectively, for the $z - Q^2$ projections and from about 1.19 GeV$^2$ to 10.24 GeV$^2$ and 0.034 to 0.45, respectively, for the $z - x$ projections, with $0.2 \leq z \leq 0.8$.
In addition, we include for the first time multiplicity
data for $K^{\pm}$ production from the Compass Collaboration~\cite{COMPASS}, which are given as a function of $z$ in bins of
inelasticity $y$ and the initial-state momentum fraction $x$.
Experimental information is available for $0.004 \leq x \leq 0.7$
and $1.2 \leq Q^2 \leq 60 $ GeV$^2$.

Finally, we update and add new sets of data for inclusive high-$p_T$ kaon production in $pp$ collisions with respect
to those included in the DSS 07 analysis. Most noteworthy are the first results for the kaon-to-pion ratio from
the ALICE Collaboration~\cite{Alice} at CERN-LHC, covering unprecedented c.m.s. energies of up to 2.76 TeV. In addition,
we include Star data taken at $\sqrt{s}=200$ GeV for charged kaon production and for the $K^-/K^+$ ratio from the Star Collaboration~\cite{Star}.
As was discussed in detail in Ref.~\cite{DSS14} in the context of pion FFs, it turns out that a good global fit of RHIC and
LHC $pp$ data, along with all the other world data, can only be achieved if one imposes a cut on the minimum
transverse momentum of the produced hadron of about 5 GeV. We maintain this cut also for the present global analysis. Such a $p_T-$cut eliminates some old $pp$ data sets included in the previous DSS 07 analysis from the fit.

\section{Results}
\subsection{Parton-to-kaon fragmentation functions}
The new set of parton-to-kaon FFs at NLO evolved at $Q^2=10$ GeV$^2$ is shown as a funcion of $z$ in Figure~1.
\begin{figure}[htb!]
\begin{center}
\includegraphics[width=0.65\textwidth]{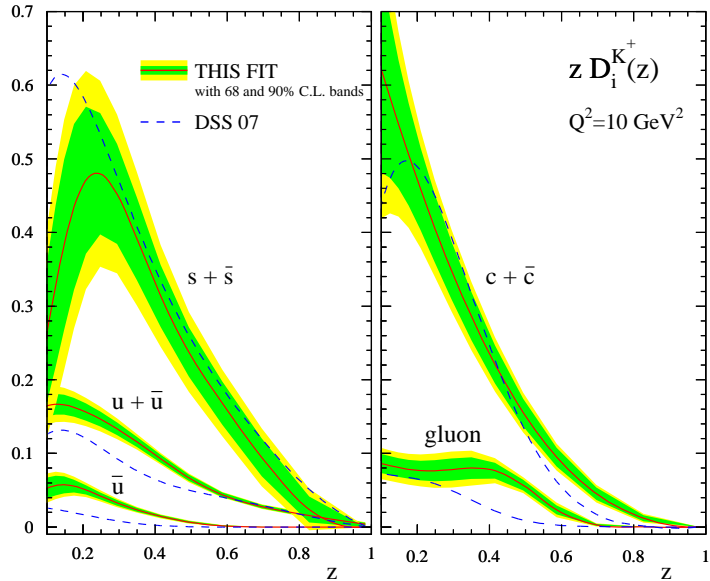} 
\caption{ 
The individual FFs for positively charged kaons $z D^{K^+}_i (z, Q^2)$ at $Q^2 = 10$ GeV$^2$.
}
\end{center}
\end{figure}
The differences with respect to the DSS 07 results are mainly driven by the newly added Belle and Babar data at high $z$, by the $z - x$ projections of the multiplicities both from Hermes and Compass, and by the $K^-/K^+$ ratios measured in pp collisions by Star and the recent multiplicities released by the ALICE Collaboration. A single parameterization for
$
D^{K^+}_u = D^{K^+}_d = D^{K^+}_d = D^{K^+}_s
$
is still the most economical choice to reproduce the data, as was the case in the original DSS 07 analysis. 
The overall quality of the fit is summarised in Table 1.
\begin{table}[bth!]
\begin{center}
\begin{tabular}{|lcccc|}
\hline 
\hline
experiment&  &  & \# data in fit & $\chi^2$ 
         \\\hline
{\sc Tpc} \cite{TPC}  &             &   & 12 & 13.4 \\
{\sc Sld} \cite{SLD}  &   &   & 48 & 81.9 \\
{\sc Aleph} \cite{ALEPH}    & & & 13 &  29.7 \\
{\sc Delphi} \cite{DELPHI}  &   &  & 36 & 31.0 \\
{\sc Opal} \cite{OPAL}  & & & 25 & 100.1 \\
{\sc BaBar} \cite{Babar}     & & & 45  & 30.6 \\ 
{\sc Belle} \cite{Belle}         & & & 78  & 15.6 \\    \hline 
{\sc Hermes} \cite{Hermes}  & & & 288 & 389.3 \\
{\sc Compass} \cite{COMPASS} & &       & 618   &  550.9     \\ \hline
{\sc Star} \cite{Star}    
                                             &                         &  &  16  & 7.6        \\    
{\sc ALICE} \cite{Alice} &                        &  &  15 & 21.6        \\ \hline\hline
{\bf TOTAL:} & & & 1194 & 1271.7 \\
\hline
\end{tabular}
\caption{\label{tab:exppiontab}Data sets used in our NLO global analysis.}
\end{center}
\end{table}
It is also worth mentioning that there is a more than five-fold increase in the number of available data points as compared to the original DSS 07 analysis. Secondly, the overall quality of the global fit has improved dramatically from $\chi^2/d.o.f. = 1.83$  for DSS 07 to $\chi^2/d.o.f. = 1.08$ for the current fit. The biggest improvement concerns the SIDIS multiplicities from Hermes which are described rather well by the updated
fit. Also, the charged kaon multiplicities from Compass and the new SIA data from BaBar and Belle integrate very nicely into the global QCD analysis. In the present fit, the set of PDFs has have been upgraded to the recent MMHT 2014 analysis, that gives a much more accurate description of sea-quark parton densities on which the analysis of SIDIS multiplicities depends strongly. It is worth mentioning that a better $\chi^2$ have found recently when MMHT 14 is replaced by NNPDF3.0~\cite{NNPDF}.
\begin{figure}[htb]
\begin{center}
\includegraphics[width=0.72\textwidth]{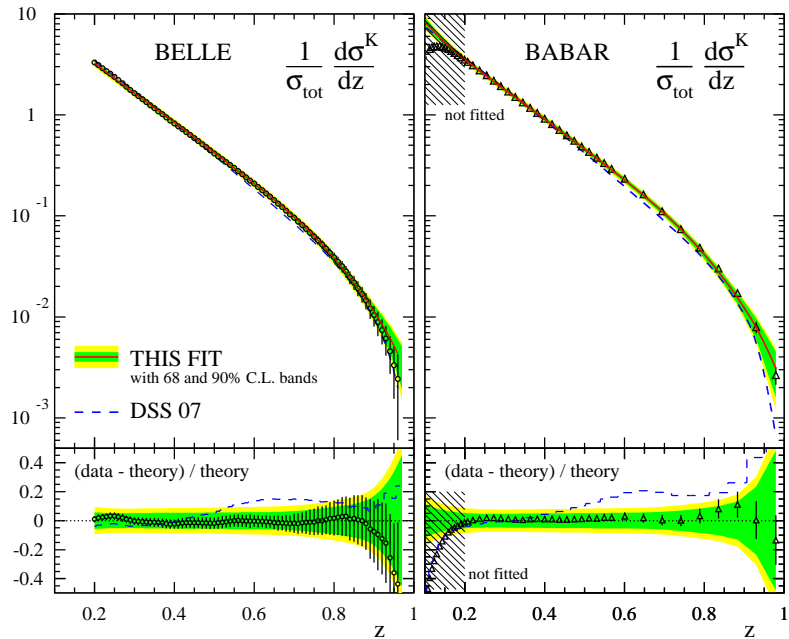} 
\caption{Results of the global QCD fit at NLO for SIA data points. On the left panel the results for Belle while in the right panel the results for BaBar.}
\end{center}
\end{figure}
\begin{figure}[htb]
\begin{center}
\includegraphics[width=0.85\textwidth]{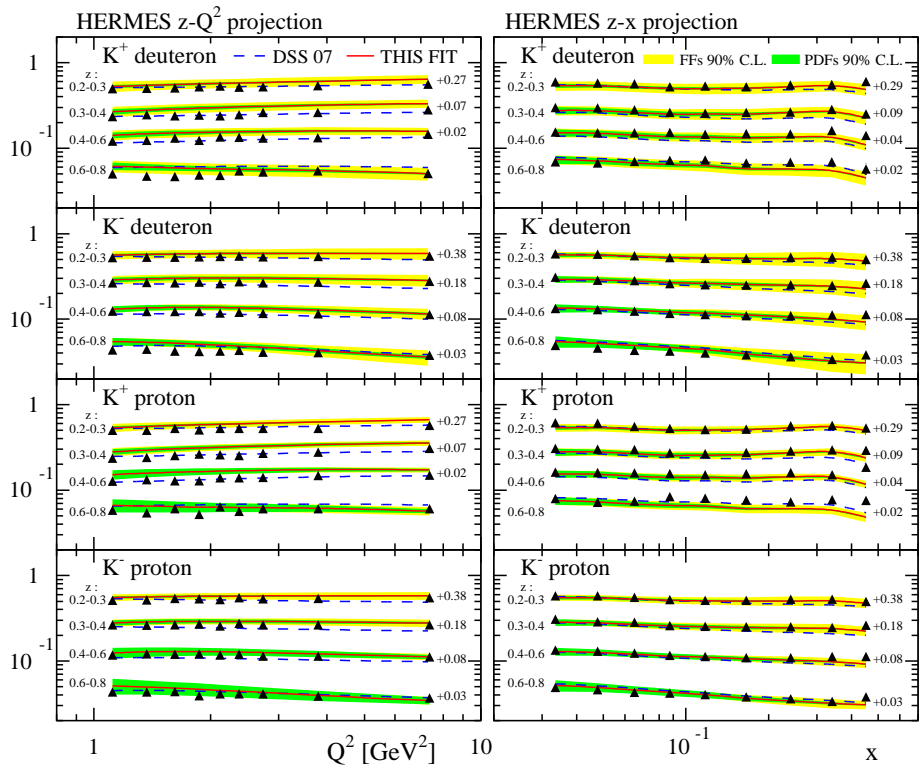} 
\caption{Results of the global QCD fit at NLO for SIDIS data points. On the left panel the results for the projection $z-Q^2$ and the right panel presents the projection $z-x$ both for the Hermes data.}
\end{center}
\end{figure}
\begin{figure}[htb]
\begin{center}
\includegraphics[width=0.60\textwidth]{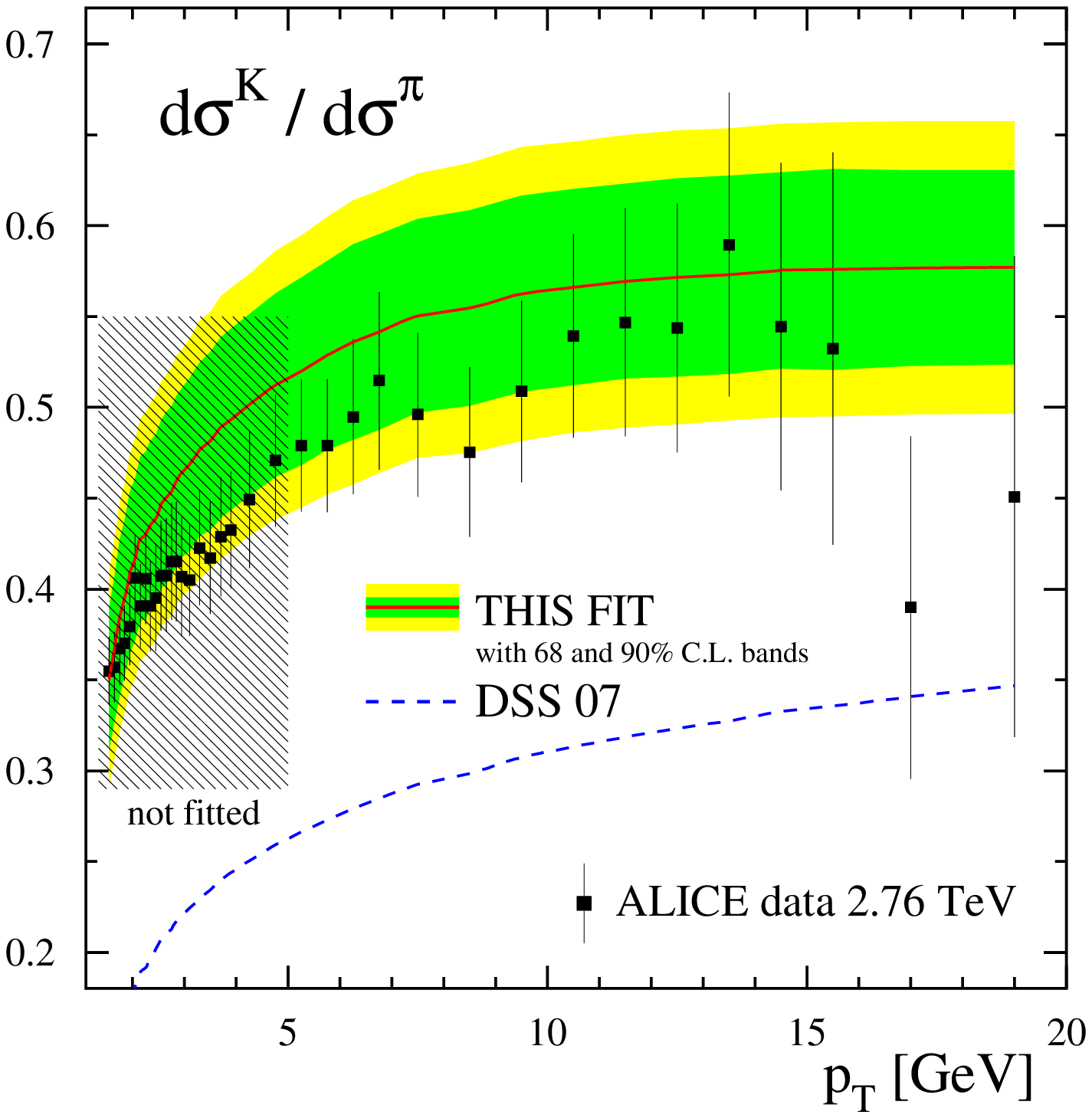}
\caption{Results of the global QCD fit at NLO for ALICE data points.}
\end{center}
\end{figure}

\subsection{Experimental data and the global QCD fit}
In this short review, we only comment on the data provided by the Belle and BaBar experiments, the multiplicities from Hermes and the measurements from the ALICE experiment. However, the reader is encouraged to read the complete analysis in Ref.~\cite{DSS17}.

In Figure 2, 3 and 4, we present a comparison of the result of the fit  at 68\% and 90\% C.L. with the SIA, SIDIS and proton-proton collisions. Regarding to Figure 2, the agreement of the fit is excellent in the entire energy and $z$-range covered by Belle and BaBar experiments. The new fit improves very significantly the description of the newly added Belle and BaBar data as can be best seen from the ``(data-theory)/theory". The estimated uncertainty bands at 68\% and 90\% C.L. reflect the accuracy and kinematical coverage of the fitted and data sets and, hence, increase towards both small and large values of $z$.

Turning into the SIDIS data, in Figure~3 we plotted the multiplicities $M_{\ell, p (d)}^{K^{\pm}}$ which are defined as the ratio of the inclusive kaon yield and the total DIS cross section in the same $x$ and $Q^2$ bin in the lepton-deuteron scattering:
\begin{eqnarray}
M_{\ell, p (d)}^{K^{\pm}}=\frac{d\sigma_{\ell, p (d)}^{K^{\pm}}/dx\, dQ^2\, dz}{d\sigma_{\ell, p (d)}/dx\, dQ^2}.
\end{eqnarray}
In the global fit we consider the two-dimensional projections of the three-dimensional multiplicity data ontho the $z-Q^2$ dependence and also the $z-x$ dependence.
A good agreement among all data points, demonstrate that the low-energy Hermes charged kaon multiplicity data can be described correctly and, equally important, without spoiling the agreement with SIA results. 

The third and final main ingredient in our global analysis is the experimental information coming from hadron-hadron collisions. 
In Figure~4, we show the charged kaon to charged pion cross section ratio as a function of the transverse momentum $p_T$ as measured by the ALICE collaboration in $pp$ collisions at midrapidity at a c.m.s. energy of 2.76 TeV. As can be seen, the current description of the data is much better than the one achieved bt the previous DSS 07 sets of pion and kaon FFs which turns out to be way too small in the entire range of $p_T$. One reason is the much reduced gluon-to-pion FF in the DSS 14 set as compared to DSS 07, which pushes the kaon-to-pion ratio up.

\section{Conclusions}

We have presented a new, comprehensive global QCD analysis of parton-to-kaon fragmentation functions
at NLO accuracy including the latest experimental information. 
The very satisfactory and simultaneous description of all data sets within the estimated uncertainties strongly supports
the validity of the underlying theoretical framework based on pQCD and, in particular, the notion of factorization
and universality for parton-to-kaon fragmentation functions.

\section{Acknowledgments}
The work of R. J. H. P. is supported by CONACyT and PROFAPI 2015 Grant No. 121. This work is also supported in part by CONICET, ANPCyT, and the Institutional Strategy of the University of T\"ubingen (DFG, ZUK 63).

\end{document}